\documentclass[12pt]{article}
\usepackage[T1]{fontenc}
\usepackage[utf8]{inputenc}
\newcommand{\R}   {{\mbox{R\hskip-0.9em{}I \ }}}
\newcommand{\N}   {{\mbox{N\hskip-0.9em{}I \ }}}

\usepackage{amsfonts}
\usepackage{amsmath}
\usepackage{amssymb}
\usepackage{mathrsfs}
\usepackage{latexsym}
\usepackage{multicol}
\usepackage{fancybox}
\usepackage{dsfont}
\usepackage{color}
\usepackage{hyperref}
\usepackage{subfigure} 
\usepackage{graphicx} 
\usepackage{caption}
\def\be{\begin{equation}}
\def\ee{\end{equation}}

\def\d'{``}

\newtheorem{thm}{Theorem}[section]

\newtheorem{propn}[thm]{Proposition}

\def\be{\begin{equation}}
\def\ee{\end{equation}}
\def\bea{\begin{eqnarray}}
\def\eea{\end{eqnarray}}

\def\i'{\textrm{i}}

\begin{document}

\begin{center}
\Large{\bf{~The Gross-Pitaevskii equation: B\"acklund  transformations and admitted solutions.}}
\end{center}

\begin{center}
{ \large{Sandra Carillo\footnote{Dipartimento Scienze di Base e Applicate per l'Ingegneria, 
\textsc{Sapienza}  Universit\`a di Roma, ROME, Italy \&
 I.N.F.N. - Sezione Roma1
Gr. IV - Mathematical Methods in NonLinear Physics.
} and Federico Zullo\footnote{DICATAM, Universit\'a di Brescia, Brescia, Italy.}  }}

{ }

\end{center}

\bigskip

 \centerline{\it In honour of Tommaso Ruggeri on his 70th Birthday.}
 \medskip 

\medskip
\medskip

\begin{abstract}
\noindent
B\"acklund  transformations are applied to study the Gross-Pitaevskii equation. Supported by previous results,  a 
class of B\"acklund transformations admitted by  this equation are constructed. 
Schwarzian derivative as well as its invariance properties turn out to represent
a key tool in the present investigation.
Examples and explicit solutions of the Gross-Pitaevskii equation are obtained. 
\end{abstract}

\bigskip\bigskip

\noindent

\noindent
{\bf Keywords}: nonlinear ordinary differential equations, Gross-Pitaevskii equation, 
B\"acklund transformations, Schwarzian derivative.

\section{Introduction} \label{intro}
The present investigation refers to B\"acklund transformations and their applications 
to the study of nonlinear differential equations. A wide literature testifies the importance of 
B\"acklund transformations in  connections to both  properties of differential equations and their 
geometrical interpretation, as well as construction of solutions to nonlinear problems. The subject is 
illustrated in the recents books   \cite{RogersSchief,Gu-book} wherein many reference are provided. 
The present study is concerned about the study of nonlinear ordinary differential equations \cite{Rogers0, Rogers1, Rogers2, CZ} via applications of  B\"acklund transformations. Indeed, they represent a key tool to reveal connections between different solutions admitted by the same nonlinear ordinary differential equation or to link solutions of different nonlinear ordinary differential equations and, therefore, in the first case,  are termed auto-B\"acklund transformations.
A wide variety of interesting results based on  B\"acklund transformations 
concern nonlinear evolution equations: they allow to investigate their structural properties, such as the 
Hamiltonian and/or bi-Hamiltonian structure and also to find solutions as well as enjoyed invariances.

In  \cite{CZ}, extending the results obtained in \cite{JMAA2000}, we constructed an algebraic map among solutions of a class of nonlinear second order differential equations. In particular, the study concerns equations of the form
\begin{equation}\label{eq}
y y''=F(z,y^2),
\end{equation}
where, as usual, the prime sign denotes derivative  with respect to the independent 
variable and $F$ is a suitably regular given function of its arguments. In  \cite{CZ}  invariance 
properties admitted by the Schwarzian derivative play a key role in the construction of 
 B\"acklund transformations;  given smooth enough function $f(z)$, its Schwarzian derivative 
 $\{f,z\}$ is defined via
\begin{equation}\label{Schw}
\{f,z\}:=\frac{f'''}{f'}-\frac{3}{2}\left(\frac{f''}{f'}\right)^2.
\end{equation}

According to \cite{CZ}, whenever  the function $F(z,v)$ satisfies {\it suitable} conditions, 
then, given any solution $y_0(z)$ of equation (\ref{eq}), another solution of same equation is 
represented by
\begin{equation}\label{bt}
y_1(z)^2=\frac{y_0(f(z))^2}{f'(z)}.
\end{equation}
Specifically, the  function   $F(z,v)$ is required to admit a  (formal) series expansion of the form 
\begin{equation}\label{F}
F(z,v)=\sum_n a_n \left(G'_n (z)\right)^{n+1}v^n-\frac{1}{2}\{w(z),z\}v,
\end{equation}
where the coefficients, given in terms of the function $G_n(z)$ and $w(z)$, satisfy the functional equations 
\begin{equation}\label{functionaleqs}
G_n(f(z))=G_n(z)+K_n, \quad w(f(z))=\frac{aw(z)+b}{cw(z)+d},~\forall a,b,c,d \qquad ad-bc\neq 0.
\end{equation}  

A large number of physically relevant differential equations belong to 
the general  class of functions (\ref{F}):
  Ermakov-Pinney equation and    the Emden-Fowler equation are shown to belong to 
  this class \cite{CZ}. 

The Ermakov-Pinney equation was introduced in 1880 by Ermakov \cite{Ermakov}, who investigated
solvability conditions of  second order ordinary differential equations. It is closely connected to the harmonic oscillator 
with a time-dependent frequency, but has applications in many other fields: for example, it describes  the motion of charged particles in the Paul trap \cite{MGW}, {{is applied}} in atomic 
transport  theory  \cite{T}, Bose-Einstein condensate theory \cite{Nicolin} and cosmolgy \cite{HL}. Furthermore,
 it plays a fundamental role in the description of the unitary evolution of quantum non-
 autonomous systems \cite{VV}. 

The Emden-Fowler equation finds applications in a wide variety of problems: for example, 
it describes the hydrostatic and thermodynamic equilibrium 
of stars, it appears in the dynamics of the Einstein's field equations \cite{Hertl}, 
in the mean-field description of critical adsorption \cite{GR} 
(see \cite{GH} and references therein for further example). Furthermore, the Emden-Fowler 
equation appears in the description of the  spherically symmetric steady 
state solutions of evolution problems involving the Laplace operator in a 
$n$-dimensional space.

The present study concernes an application of the transformations (\ref{bt}) to the Gross-Pitaevskii 
equation. The Gross-Pitaevskii equation is a non-linear Schr\"odinger equation with a cubic 
non-linearity: it describes the dynamics of Bose-Einstein condensates. 
Other interesting phenomena, such as  the propagation of electromagnetic waves in optical 
fibers with variable refractive index are also  described via the Gross-Pitaevskii equation. 
\medskip

The article is organised as follows: in Section \ref{secbt}, recalling  results in \cite{CZ},  
auto-B\"acklund transformations admitted by equation (\ref{eq}) are obtained.
Notably, the Schwar\-zian derivative and a functional-differential equation satisfied by the 
function $F(z,v)$  in (\ref{eq}) play a crucial role to obtain the result. In Section \ref{secres} 
we reduce the Gross-Pitaevskii  equation  and equation (\ref{eq}) is specialised to give the 
reduced case under consideration. 
                               
\section{The B\"acklund transformations.}\label{secbt}
In this section, for completeness, some of the results given in \cite{CZ} are briefly overviewed. We wish to recall only the main propositions: the interested reader may find  {{the}} details in the original reference. First of all,  notice that if we let $v=y^2$, equation (\ref{eq}) is equivalent to the following equation \begin{equation}\label{2}
v'' v -\frac{1}{2}(v')^2=2v F(z, v),
\end{equation} 
wherein the unknown is $v$. The subsequent application of  the reciprocal transformation
\begin{equation}\label{reciprocal}
R:  ~~~~ \left\{\begin{array}{ccc}x = \Phi &  {D_t} = v {D_x},  \\ \\
  v =  D_t \Phi  & ~~~~~ D_x = [ \Phi_t]^{-1}{D_t}  \end{array}\right.
  \end{equation}
where
\begin{equation}\label{R33}
\displaystyle{ {D_t} := {d  \over
dt},~~~~ {D_x} :=  {d  \over dx}, ~~~ \Phi_t = {d  \over dt} \Phi},
\end{equation}
shows that equation \eqref{eq} is linked to the Schwarzian equation
\begin{equation}\label{eqphi}
\{\phi,t\}=2\phi_t F(\phi,\phi_t).
\end{equation}
The latter,   enjoys interesting {{invariance properties}}; in particular, it is invariant under any transformation of the form  $\phi(t)=g(\psi(t))$, where  $g(\cdot)$ is a suitably regular function of its argument. Namely, if  $\phi(t)$ is regarded as a composed function of $t$, i.e. if  $\phi$  can be expressed as 
$\phi(t)=g(\psi(t))$, then $\psi$ satisfies a Schwarzian equation of the same form of  (\ref{eqphi}) 
\begin{equation}\label{eqphi2}
\{\psi,t\}=2\psi_t \tilde F(\psi,\psi_t)~,
\end{equation}
where $\tilde{F}(\psi, \psi_t)$ is explicitly given by (see \cite{CZ}) 
\begin{equation}
\tilde{F}(\psi,\psi_t):= g'F(g,g'\psi_t)-\frac{\psi_t}{2}\{g,\psi\}.
\end{equation}
It follows that if $y(z)$ is a solution of equation (\ref{eq}) and if we introduce a function $Y(z)$ related to $y(z)$ by
\begin{equation}\label{map}
Y^2(z)f'(z)-y^2(f(z))=0,
\end{equation}
then $Y(z)$ is a solution of the equation
\begin{equation}\label{te}
YY''=f'F(f,f'Y^2)-\frac{1}{2}\{f,z\}Y^2~.
\end{equation}
The transformation (\ref{map}) maps a solution of equation (\ref{eq}) to a solution of equation (\ref{te}): it is a B\"acklund transformation between the two equations.

The previous results imply, however, that if the function $F(z,v)$ satisfies the following functional-differential equation
\begin{equation}\label{fde}
F(z,v)=f'F(f,f'v)-\frac{1}{2}\{f,z\}v,
\end{equation}
then the transformation (\ref{map}) maps a solution of equation (\ref{eq}) to a solution of the \emph{same} equation: in this case we get an auto-B\"acklund transformation admitted by  equation (\ref{eq}). In general it is very hard to characterize the solutions of a functional-differential equation, so it could seem that the identification and description of the class of  auto-B\"acklund transformation of the form (\ref{map}) for equation (\ref{eq}) is hopeless. However, as shown in \cite{CZ}, if one posit a formal power expansion in $v$ for the function $F(z,v)$, i.e. if one assumes that $F(z,v)$ can be represented via a formal series expansion of the form
\begin{equation}\label{Fexp}
F(z,v)=\sum_{n}F_{n}(z)v^n, ~~n\in K\subset\N
\end{equation} 
then some interesting structural properties  of equation (\ref{fde}) can be understood. Indeed, we may 
consider (\ref{fde})  as a linear non homogeneous ordinary differential equation in the unknown  $F$.
 This approach allows us to say which forms the given  function $F$ may assume in order  to be 
 amenable to the presented method. Specifically, given a suitable $F$, a method to construct 
 solutions of equation \eqref{eq}, via  B\"acklund transformations, can be provided.

For the sake of convenience, when $n\neq 1$, the 
coefficients $F_n(z)$ are looked for under the form $F_{n}=(G'_n)^{n+1}$, where $G'_n$ denotes the 
derivative of a smooth (at least $C^1$) function of $z$. In the case $n=1$, we set 
\begin{equation}\label{F1}
F_1(z):=(G'_1)^{2}-\frac{1}{2}Q(z)
\end{equation}
where $Q(z)$ is a suitable function. Hence, substitution of the latter and of  \eqref{Fexp} in (\ref{fde}),  gives 
\begin{equation}\label{fde1}
2\sum_{n}\left((f')(G'_n)(f)-(G'_n)(z)\right)^{n+1}v^n-(f')^{2}Q(f)v+Q(z)v-\{f,z\}v=0.
\end{equation}
which depends explicitly on $G_n$, so that, for example, $(G'_n)(f)$ indicates that  $G'_n$ is composed with $f(z)$.
 \\
The power series expansion in (\ref{fde1}) represents a polynomial in $v$ whose first term multiplies $v^{-1}$,  hence equality (\ref{fde1}) is satisfied when all the coefficients of  $v^{k}$ are set 
equal to zero. This means that the action of the map $f(z)$ on the functions $G_n(z)$ is a translation, 
that is, the functions $G_n(z)$ satisfy the linear functional equations
\begin{equation}\label{G_nK}
G_n(f(z))=G_n(z)+K_n~~~\Longleftrightarrow~f'G'_n(f)=G'_n(z)~~\forall n,
\end{equation}
where $K_n$ are arbitrary constants.  Substitution of \eqref{G_nK} in (\ref{fde1}) gives
\begin{equation}\label{fde2}
Q(z)=(f')^{2}Q(f)+\{f,z\}.
\end{equation}
This equation suggests to identify $Q$ with a suitable Schwarzian derivative, that is, to set
\begin{equation}\label{position}
Q(z):=\{w,z\}.
\end{equation}
Substitution of this position in \eqref{G_nK} shows that solutions of \eqref{fde2}  are obtained whenever  $w(z)$ and $f(z)$ are related via the functional equation
\begin{equation}\label{position2}  
\{w(z),z\}=(f')^{2}\{w(f),f\}+\{f,z\}, 
\end{equation}
wherein the Schwarzian derivative of the composition of two functions appears since 
$(f')^{2}\{w(f),f\}+\{f,z\}\equiv \{w(f),z\}$. Accordingly, the function $w(f(z))$ and $w(z)$ follow to be related via a fractional linear transformation, i.e.
\begin{equation}\label{flt}
w(f(z))=\frac{aw(z)+b}{cw(z)+d}~~, \qquad ad-bc\neq 0.
\end{equation}
Hence, combination of the shown results, allows to prove the claims made in the Introduction, summarized in the following two propositions (see also \cite{CZ}).
\begin{propn}\label{prop1}
If the function $F(z,v)$ can be represented as a power series expansion in $v$, then the solution 
of the functional differential equation
\begin{equation*}
F(z,v)=f'F(f,f'v)-\frac{1}{2}\{f,z\}v.
\end{equation*}
is given by
\begin{equation}\label{F2}
F(z,v)=\sum_n a_n \left(G'_n (z)\right)^{n+1}v^n-\frac{1}{2}\{w(z),z\}v
\end{equation}
where the functions $G_n(z)$ and $w(z)$ satisfy the functional equations ($a$, $b$, $c$, $d$ and $K_n$ are arbitrary constants)
\begin{equation}\label{functionaleqs1}
G_n(f(z))=G_n(z)+K_n, \quad w(f(z))=\frac{aw(z)+b}{cw(z)+d}, \qquad ad-bc\neq 0.
\end{equation}
\end{propn}
\smallskip
\begin{propn}\label{cor1}
Let $y_0(z)$ be a solution of the differential equation
\begin{equation}\label{diffeq}
y''=\sum_n a_n \left(G'_n (z)\right)^{n+1}y^{2n-1}-\frac{1}{2}\{w(z),z\}y,
\end{equation}
where the functions $G_n(z)$ and $w(z)$ are specified in Proposition \ref{prop1}, then 
$$
y_1(z)^2=\frac{y_0(f(z))^2}{f'(z)}
$$
represents an auto-B\"acklund transformation admitted by equation (\ref{diffeq}). 
\end{propn}

\section{The Gross-Pitaevskii equation and  B\"acklund transformations.}\label{secres}
The non-linear Schr\"odinger equation with a potential depending on the spaces coordinates is often called the Gross-Pitaevskii equation. The equation has a cubic non-linearity and plays a special role in mathematical physics: it is widely applied in the description of light propagation   inside an optical fiber \cite{KA} and in Bose-Einstein condensates theory \cite{Tor}. The cubic term, in the last case,  describes  particles interactions: it can be modulated by a function depending on the space coordinates. When, as usual, the external potential (e.g. magnetic potential trapping the particles) is described by a function $V(x)$ and  all the quantities are expressed in natural units, the equation takes the form (see e.eg. \cite{Tor})
\begin{equation}\label{GP}
\textrm{i}\frac{\partial \psi}{\partial t}=-\frac{1}{2}\frac{\partial^2 \psi}{\partial x^2}+V(x)\psi+g(x)|\psi|^2\psi. 
\end{equation}
The ansatz
$$
\psi(x,t)=r(x)e^{\textrm{i}\left(\theta(x)-\mu t\right)}
$$
where $r$, $\theta$ and $\mu$ are all real, allows to split equation (\ref{GP}) into its real and imaginary part as
$$
r\theta '' + 2r'\theta'=0,\qquad  r''-2gr^3-\left((\theta')^2+2V-2\mu\right)r=0.
$$ 
The first equation is readily solved by
$$
r^2\theta'=c, ~~~\forall c\in\R.
$$
Substitution  of $\theta'$ in the second equation allows to obtain the following non-linear ODE
\begin{equation}\label{GP1}
r(x)''=\frac{c^2}{r(x)^3}+2(V(x)-\mu)r(x)+2g(x)r(x)^3.
\end{equation}
let us now consider equation (\ref{GP1}) in the special case when the potential $V(x)$ and the interaction term $g(x)$ can be represented by the following formulae:
$$
2(V(x)-\mu)=-\frac{1}{2}\{w,x\},\qquad 2g(x)=b(G(x)')^3
$$
where the functions $G(x)$ and $w(x)$ satisfy the functional equations (\ref{functionaleqs}). Then the results of the previous section apply (see propositions (\ref{prop1}) and (\ref{cor1})) and, if $r_0$ is a solution of equation (\ref{GP1}), then another solution $r_1(x)$  follows from the  auto-B\"acklund transformation
$$
r_1(x)^2f'(x)-r_0(f(x))^2.
$$
As an example let $G(x)$ be a polynomial, and, specifically, let 
\begin{equation}
G(x)=x^n(1+\eta x^n ) ~~,~ n\in \N, ~  \eta\in\R.
\end{equation}
Then, equation (\ref{GP1}) is given by
\begin{equation}\label{GPexp}
r(x)''=\frac{c^2}{r(x)^3}+\left(\frac{n^2-1}{4x^2}+\frac{3x^{2n-2}\eta^2n^2}{(1+2\eta x^n)^2}\right)r(x)+bx^{3n-3}(1+2\eta x^n)^3r(x)^3.
\end{equation}
Let  $f(x)$ denote a solution of the polynomial equation
$$
f(x)^n(1+\eta f(x)^n )=x^n(1+\eta x^n )+K,
$$
which if $\eta K$ is greater than $\frac{1}{4}$, admits always  a real positive root. 
If $r_0(x)$ is a solution of equation (\ref{GPexp}), then another solution is given by
\begin{equation}\label{btex}
r_1(x)^2=r_0(f(x))^2\frac{f(x)^{n-1}(1+2\eta f(x)^n)}{x^{n-1}(1+2\eta x^{n})}~.
\end{equation}
The latter shows that a particular solution of equation (\ref{GPexp}) can be found by setting $r(x)^2x^{n-1}(1+2\eta x^{n})$ equal to a positive constant. Indeed it is possible to check that the function
\begin{equation}\label{r}
r(x)=\frac{v}{\sqrt{x^{n-1}(1+2\eta x^{n})}}
\end{equation}
is a solution of equation (\ref{GPexp}) provided that the condition $bv^6+c^2=0$, constraining the value of $b$, is satisfied. The previous solution is a fixed point of the transformation (\ref{btex}). The solution for the wave function $\psi(x,t)$ corresponding to (\ref{r})  reads
$$
\psi(x,t)=v\frac{\exp\left(\textrm{i}\left(\frac{c^2x^n(1+\eta x^n)}{n v^2}+\theta(0)-\mu t\right)\right)}{\sqrt{x^{n-1}(1+2\eta x^{n-1})}}
$$
The previous solution is bounded at the origin provided  $n\leq1$.
\section{Conclusions}

The results here presented are part of a long time investigation program which covers various aspects 
of the study on nonlinear evolution equations and their structural properties as well as the construction 
of  solutions admitted by particular problems (sse e.g.  \cite{PhysA166, JMP2018},  \cite{Rag0,Rag1, Rag2},  
\cite{FZ1,FZ2}). Here we applied the transformations found in \cite{CZ} to the Gross-Pitaevskii 
equation in 1+1 dimensions. Indeed the transformations found are auto- B\"acklund transformations for 
the ordinary differential equation (\ref{eq}) and  a reduction of the partial differential equation 
(\ref{GP}) {{was required}} to apply our result. The application to the Gross-Pitaevskii equation  
provided in this work can be considered a first approach to a more general research on this issue: the 
case of periodic potentials would be interesting for physical applications but also we believe that the 
analytical structures of the equations obtained deserve further efforts to fully understand the 
potentiality of the method.  In particular equation (\ref{btex}) shows that the Schr\"oder's functional 
equation $\Phi(f(x))=s\Phi(x)$ with an eigenvalue $s$ equal to 1 plays a special role in the description 
of the solutions of equation (\ref{GP}). This and other issues will be further developed in future works.

%%%%%%%%%%%%%%%%%%%%%%%%%%%%%
\begin{center} \bf Acknowledgments\end{center}
The financial  support of GNFM-INdAM, INFN, \textsc{Sapienza} Universit\`a di Roma and Unversit\`a di Brescia are gratefully acknowledged.
%%%%%%%%%%%%%%%%%%%%%%%%%%%%%

\end{document}